\documentstyle[12pt]{article}
\begin{document}
\begin{flushright}
Preprint IHEP 95-123\\
hep-ph/9602???\\
To appear in Phys.Lett.B
\end{flushright}
\begin{center}
{\large\bf Heavy quark-meson mass gap from spectroscopy}\\
\vspace*{3mm}

{V.V.Kiselev}\\
\vspace*{2mm}
Institute for High Energy Physics,\\
Protvino, Moscow Region, 142284, Russia\\
E-mail: kiselev@mx.ihep.su
\end{center}

\begin{abstract}
Using a quite accurate flavour-independence of $nS$-level mass differences
in heavy quarkonia and a corresponding quasiclassical expression for the
heavy quark binding energy, one shows that experimental data on masses of
mesons with heavy quarks allow one to make the estimate $\bar \Lambda=
0.63\pm 0.03$ GeV.
\end{abstract}

\section*{Introduction}

The QCD dynamics makes a screen in observation of pure electroweak effects in
heavy quark interactions. One of powerful tools in theoretical studies
of hadrons with a single heavy quark, is HQET \cite{1}, where the
heavy quark-meson mass gap $\bar \Lambda = m(Q\bar q) - m_Q +
O(\bar \Lambda/m_Q)$ has a significant meaning. In the $\bar \Lambda$ 
expression given above, $m(Q\bar q)$ denotes the heavy-light meson mass
$m(Q\bar q) = (3 m_{V}(Q\bar q)+m_{P}(Q\bar q))/4$, averaged over the
spin-dependent part of interaction and correspondingly expressed
through the masses of vector and pseudoscalar states, and $m_Q$ is 
the heavy quark pole mass related to its "running" mass \cite{2}. 

At present, $\bar \Lambda$ is evaluated in the following ways. The first one
is the HQET sum rules giving $\bar \Lambda = 0.5\pm 0.1$ GeV \cite{3}.
The second is the full QCD sum rules for heavy quarkonia \cite{4},
where the pole or "running" masses of heavy quarks are determined from the
experimental data on the heavy quarkonia masses and leptonic constants, so
that $\bar \Lambda= 0.6 - 0.7$ GeV \cite{5}. A special way is the
nonrelativistic version of QCD sum rules for bottomonium \cite{6},
where one can use a region of moderate values of the spectral function moment
numbers, so that the nonperturbative contribution given by the gluon condensate
can be still neglected and the Coulomb corrections can be quite accurately 
taken into account. Voloshin M. has found the "$\mu$-independent" $b$ quark
pole mass $m^*_b = m_b-0.56\alpha_s(\mu) = 4.639\pm 0.002$ GeV, corresponding
to $\bar \Lambda=0.62\pm 0.02$ GeV. Next, one considers the inclusive 
semileptonic width of a heavy-light meson in a way allowing one to take
into account hard gluon corrections to the weak current of quarks,
so that one inserts light quark loops into the gluon propagator.
The comparison of the calculated width with the experimental value
gives the "running" mass of heavy quark. Its value depends on the $l$ number
of the light quark loops. In the infinite $l$ limit, one finds $\bar \Lambda
=0.25$ GeV \cite{7}. However, the latter estimate of hard corrections is
generally based on the spectator mechanism assuming the neglecting of
the heavy quark binding into the meson. Despite the renormalon ambiguity
in the heavy quark mass determination \cite{8}, one should not 
straightforwardly conclude that there is a deep disagreement between 
the sum rule
estimates of $\bar \Lambda$ and the evaluation in the decay analysis, since
the latter is based on the consideration of isolated, but dressed heavy
quark, whereas the sum rules handle with exactly defined quantities
involving no additional assumptions like the spectator picture. Moreover,
the sum rule analysis is made for a finite number of loops.

Further, the nonrelativistic sum rule consideration of bottomonium can give
a quite accurate value of $b$ quark mass at the moment numbers, where
the gluon condensate gives a negligible contribution. In the sum rules at
high numbers of the spectral moments, the gluon condensate contribution
determines the binding energy in the $1S$-state and, hence, the
difference between the double quark mass and the level mass. However,
to use experimental data on the excited level masses and leptonic constants
is enough to extract the quark mass at the moderate values of the
spectral moment numbers. Therefore, the information on the excitation masses
can allow one to determine the heavy quark masses.

In the present paper we use the experimental regularity in the heavy 
quarkonium spectra, where one finds a quite accurate flavour-independence
in the $S$-wave level spacing. This approximate independence results 
in a number of explicit relations between the quark masses, excitation 
numbers and binding energies \cite{9}. These equations compose a complete
system, which allows one to determine the heavy quark-meson mass gap 
of high accuracy
$$
\bar \Lambda = 0.63\pm 0.03\; \mbox{GeV.}
$$

\section{Basic relations}

Determine the heavy quark pole mass
\begin{equation}
m_Q = m(Q\bar q) -\bar \Lambda - \frac{\mu^2_\pi}{2m(Q\bar q)}
+O(1/m^2_Q)\;,
\end{equation}
where $\mu^2_\pi$ is the average square of quark momentum inside a 
heavy-light meson. In the following we put
\begin{equation}
\mu^2_\pi = 2\langle T\rangle \mu_{Q\bar q}\;, \label{2}
\end{equation}
where $\mu_{Q\bar q}$ is the reduced mass of the $Q\bar q$ system, and
$T$ is the kinetic energy of quarks. 
The reasonable choice of $\mu_{Q\bar q}$ is $\bar \Lambda$ \cite{1,9+}.

The $m(Q\bar q)$ values for $Q=b,\; c$ and $q=u,\; d$ are known 
experimentally \cite{10}
\begin{equation}
m_B(1S) =5.313\; \mbox{GeV,}\;\;\; m_D(1S) = 1.975\; \mbox{GeV,}
\end{equation}
with the accuracy better than 5 MeV.

By the Feynman - Hellmann theorem for the $Q_1\bar Q_2$ system
\begin{equation}
\frac{dE}{d\mu_{12}} = - \frac{\langle T\rangle}{\mu_{12}}\;,
\;\;\; \mu_{12} = \frac{m_1m_2}{m_1+m_2}\;, \label{3}
\end{equation}
the flavour-independent spacing of heavy quarkonium levels is reached
if $\langle T\rangle=T$ is the constant value. Relation (\ref{3})
fixes the quark mass dependence of the binding energy in heavy quarkonia.
To complete the system of equations, we use the Bohr - Sommerfeld
quantization of $nS$-states at the constant $\langle T\rangle$ in the
flavour-independent logarithmic potential
\begin{equation}
E(n) = C + T\ln{\frac{n^2}{\mu_{12}}}\;, \label{4}
\end{equation}
where $C$ is a flavour-independent constant. Eq.(\ref{4}) gives
the excitation number dependence of the binding energy in the heavy quarkonia.
Note, that the semiclassical WKB approximation of 3-dimensional potential 
problem leads to the substitution of $n-1/4$ for $n$ in (\ref{4}).
However, the latter substitution does not result in the better description of
experimental data\footnote{The same tendency was certainly observed
in the Coulomb potential, where the Bohr--Sommerfeld equation gave
exact results.} (Fig.1). To isolate the $n$ dependence, we consider
the ratio $\alpha(n)$
\begin{equation}
\alpha(n) = \frac{M(nS) - M(1S)}{M(2S) - M(1S)} = \frac{\ln n}{\ln 2}
\label{5}
\end{equation}
in the current model or
$$
\alpha^{\rm WKB}(n) = \frac{\ln[(4n-1)/3]}{\ln[7/3]}\;.
$$
The comparison of model approximation (\ref{5}) with the experimental 
values \cite{10} and WKB modification is shown in Fig.1. One can see 
that the applied model is more suitable for the accurate description of
$M(nS)$ values with the accuracy up to 30 MeV, so that the parameter
$T$ is equal to
\begin{equation}
T = 0.43\pm 0.02\; \mbox{GeV.}
\end{equation}
Note, that in contrast to the analysis in \cite{9}, we use the excitation
masses averaged over the spin-dependent part of a potential, this procedure 
is more reasonable. In addition, the analysis of heavy quarkonium spectra
performed in \cite{9} under the WKB approximation results in the smaller 
value of $T\approx 0.37$ GeV.
\setlength{\unitlength}{0.85mm}\thicklines
\begin{figure}[t]
\begin{center}
\begin{picture}(100,90)
\put(15,10){\framebox(60,70)}
\put(3,10){$0$}
\put(15,30){\line(1,0){2}}
\put(3,30){$1$}
\put(15,50){\line(1,0){2}}
\put(3,50){$2$}
\put(15,70){\line(1,0){2}}
\put(3,70){$3$}

\put(15,20){\line(1,0){2}}
\put(15,40){\line(1,0){2}}
\put(15,60){\line(1,0){2}}

\put(0,83){$\alpha(n)$}

\put(25,10){\line(0,1){2}}
\put(35,10){\line(0,1){2}}
\put(45,10){\line(0,1){2}}
\put(55,10){\line(0,1){2}}
\put(65,10){\line(0,1){2}}
\put(15,2){$1$}
\put(25,2){$2$}
\put(35,2){$3$}
\put(45,2){$4$}
\put(55,2){$5$}
\put(65,2){$6$}
\put(75,2){$n$}

\put(15,10){\circle*{1.6}}
\put(25,30){\circle*{1.6}}
\put(35,41.8){\circle*{1.6}}
\put(45,49.8){\circle*{1.6}}
\put(55,60){\circle*{1.6}}
\put(65,65.6){\circle*{1.6}}

\put(14,09){\framebox(2,2)}
\put(24,29){\framebox(2,2)}
\put(34,41){\framebox(2,2)}
\put(44,45){\framebox(2,2)}
\put(54,53.8){\framebox(2,2)}

 \put( 15.10, 10.29){\circle*{0.5}}
 \put( 15.10, 10.31){\circle*{0.4}}
 \put( 15.20, 10.57){\circle*{0.5}}
 \put( 15.20, 10.62){\circle*{0.4}}
 \put( 15.30, 10.85){\circle*{0.5}}
 \put( 15.30, 10.93){\circle*{0.4}}
 \put( 15.40, 11.13){\circle*{0.5}}
 \put( 15.40, 11.23){\circle*{0.4}}
 \put( 15.50, 11.41){\circle*{0.5}}
 \put( 15.50, 11.52){\circle*{0.4}}
 \put( 15.60, 11.68){\circle*{0.5}}
 \put( 15.60, 11.82){\circle*{0.4}}
 \put( 15.70, 11.95){\circle*{0.5}}
 \put( 15.70, 12.11){\circle*{0.4}}
 \put( 15.80, 12.22){\circle*{0.5}}
 \put( 15.80, 12.39){\circle*{0.4}}
 \put( 15.90, 12.49){\circle*{0.5}}
 \put( 15.90, 12.68){\circle*{0.4}}
 \put( 16.00, 12.75){\circle*{0.5}}
 \put( 16.00, 12.95){\circle*{0.4}}
 \put( 16.10, 13.01){\circle*{0.5}}
 \put( 16.10, 13.23){\circle*{0.4}}
 \put( 16.20, 13.27){\circle*{0.5}}
 \put( 16.20, 13.50){\circle*{0.4}}
 \put( 16.30, 13.53){\circle*{0.5}}
 \put( 16.30, 13.77){\circle*{0.4}}
 \put( 16.40, 13.78){\circle*{0.5}}
 \put( 16.40, 14.04){\circle*{0.4}}
 \put( 16.50, 14.03){\circle*{0.5}}
 \put( 16.50, 14.30){\circle*{0.4}}
 \put( 16.60, 14.28){\circle*{0.5}}
 \put( 16.60, 14.56){\circle*{0.4}}
 \put( 16.70, 14.53){\circle*{0.5}}
 \put( 16.70, 14.82){\circle*{0.4}}
 \put( 16.80, 14.78){\circle*{0.5}}
 \put( 16.80, 15.08){\circle*{0.4}}
 \put( 16.90, 15.02){\circle*{0.5}}
 \put( 16.90, 15.33){\circle*{0.4}}
 \put( 17.00, 15.26){\circle*{0.5}}
 \put( 17.00, 15.58){\circle*{0.4}}
 \put( 17.10, 15.50){\circle*{0.5}}
 \put( 17.10, 15.83){\circle*{0.4}}
 \put( 17.20, 15.74){\circle*{0.5}}
 \put( 17.20, 16.07){\circle*{0.4}}
 \put( 17.30, 15.97){\circle*{0.5}}
 \put( 17.30, 16.31){\circle*{0.4}}
 \put( 17.40, 16.21){\circle*{0.5}}
 \put( 17.40, 16.55){\circle*{0.4}}
 \put( 17.50, 16.44){\circle*{0.5}}
 \put( 17.50, 16.79){\circle*{0.4}}
 \put( 17.60, 16.67){\circle*{0.5}}
 \put( 17.60, 17.03){\circle*{0.4}}
 \put( 17.70, 16.90){\circle*{0.5}}
 \put( 17.70, 17.26){\circle*{0.4}}
 \put( 17.80, 17.12){\circle*{0.5}}
 \put( 17.80, 17.49){\circle*{0.4}}
 \put( 17.90, 17.35){\circle*{0.5}}
 \put( 17.90, 17.72){\circle*{0.4}}
 \put( 18.00, 17.57){\circle*{0.5}}
 \put( 18.00, 17.94){\circle*{0.4}}
 \put( 18.10, 17.79){\circle*{0.5}}
 \put( 18.10, 18.17){\circle*{0.4}}
 \put( 18.20, 18.01){\circle*{0.5}}
 \put( 18.20, 18.39){\circle*{0.4}}
 \put( 18.30, 18.23){\circle*{0.5}}
 \put( 18.30, 18.61){\circle*{0.4}}
 \put( 18.40, 18.44){\circle*{0.5}}
 \put( 18.40, 18.82){\circle*{0.4}}
 \put( 18.50, 18.66){\circle*{0.5}}
 \put( 18.50, 19.04){\circle*{0.4}}
 \put( 18.60, 18.87){\circle*{0.5}}
 \put( 18.60, 19.25){\circle*{0.4}}
 \put( 18.70, 19.08){\circle*{0.5}}
 \put( 18.70, 19.47){\circle*{0.4}}
 \put( 18.80, 19.29){\circle*{0.5}}
 \put( 18.80, 19.68){\circle*{0.4}}
 \put( 18.90, 19.50){\circle*{0.5}}
 \put( 18.90, 19.88){\circle*{0.4}}
 \put( 19.00, 19.71){\circle*{0.5}}
 \put( 19.00, 20.09){\circle*{0.4}}
 \put( 19.10, 19.91){\circle*{0.5}}
 \put( 19.10, 20.29){\circle*{0.4}}
 \put( 19.20, 20.12){\circle*{0.5}}
 \put( 19.20, 20.50){\circle*{0.4}}
 \put( 19.30, 20.32){\circle*{0.5}}
 \put( 19.30, 20.70){\circle*{0.4}}
 \put( 19.40, 20.52){\circle*{0.5}}
 \put( 19.40, 20.90){\circle*{0.4}}
 \put( 19.50, 20.72){\circle*{0.5}}
 \put( 19.50, 21.09){\circle*{0.4}}
 \put( 19.60, 20.92){\circle*{0.5}}
 \put( 19.60, 21.29){\circle*{0.4}}
 \put( 19.70, 21.12){\circle*{0.5}}
 \put( 19.70, 21.48){\circle*{0.4}}
 \put( 19.80, 21.31){\circle*{0.5}}
 \put( 19.80, 21.68){\circle*{0.4}}
 \put( 19.90, 21.51){\circle*{0.5}}
 \put( 19.90, 21.87){\circle*{0.4}}
 \put( 20.00, 21.70){\circle*{0.5}}
 \put( 20.00, 22.06){\circle*{0.4}}
 \put( 20.10, 21.89){\circle*{0.5}}
 \put( 20.10, 22.25){\circle*{0.4}}
 \put( 20.20, 22.08){\circle*{0.5}}
 \put( 20.20, 22.43){\circle*{0.4}}
 \put( 20.30, 22.27){\circle*{0.5}}
 \put( 20.30, 22.62){\circle*{0.4}}
 \put( 20.40, 22.46){\circle*{0.5}}
 \put( 20.40, 22.80){\circle*{0.4}}
 \put( 20.50, 22.65){\circle*{0.5}}
 \put( 20.50, 22.98){\circle*{0.4}}
 \put( 20.60, 22.83){\circle*{0.5}}
 \put( 20.60, 23.16){\circle*{0.4}}
 \put( 20.70, 23.02){\circle*{0.5}}
 \put( 20.70, 23.34){\circle*{0.4}}
 \put( 20.80, 23.20){\circle*{0.5}}
 \put( 20.80, 23.52){\circle*{0.4}}
 \put( 20.90, 23.38){\circle*{0.5}}
 \put( 20.90, 23.70){\circle*{0.4}}
 \put( 21.00, 23.56){\circle*{0.5}}
 \put( 21.00, 23.87){\circle*{0.4}}
 \put( 21.10, 23.74){\circle*{0.5}}
 \put( 21.10, 24.05){\circle*{0.4}}
 \put( 21.20, 23.92){\circle*{0.5}}
 \put( 21.20, 24.22){\circle*{0.4}}
 \put( 21.30, 24.10){\circle*{0.5}}
 \put( 21.30, 24.39){\circle*{0.4}}
 \put( 21.40, 24.27){\circle*{0.5}}
 \put( 21.40, 24.56){\circle*{0.4}}
 \put( 21.50, 24.45){\circle*{0.5}}
 \put( 21.50, 24.73){\circle*{0.4}}
 \put( 21.60, 24.62){\circle*{0.5}}
 \put( 21.60, 24.90){\circle*{0.4}}
 \put( 21.70, 24.80){\circle*{0.5}}
 \put( 21.70, 25.07){\circle*{0.4}}
 \put( 21.80, 24.97){\circle*{0.5}}
 \put( 21.80, 25.23){\circle*{0.4}}
 \put( 21.90, 25.14){\circle*{0.5}}
 \put( 21.90, 25.40){\circle*{0.4}}
 \put( 22.00, 25.31){\circle*{0.5}}
 \put( 22.00, 25.56){\circle*{0.4}}
 \put( 22.10, 25.48){\circle*{0.5}}
 \put( 22.10, 25.72){\circle*{0.4}}
 \put( 22.20, 25.65){\circle*{0.5}}
 \put( 22.20, 25.88){\circle*{0.4}}
 \put( 22.30, 25.82){\circle*{0.5}}
 \put( 22.30, 26.04){\circle*{0.4}}
 \put( 22.40, 25.98){\circle*{0.5}}
 \put( 22.40, 26.20){\circle*{0.4}}
 \put( 22.50, 26.15){\circle*{0.5}}
 \put( 22.50, 26.36){\circle*{0.4}}
 \put( 22.60, 26.31){\circle*{0.5}}
 \put( 22.60, 26.52){\circle*{0.4}}
 \put( 22.70, 26.47){\circle*{0.5}}
 \put( 22.70, 26.67){\circle*{0.4}}
 \put( 22.80, 26.64){\circle*{0.5}}
 \put( 22.80, 26.83){\circle*{0.4}}
 \put( 22.90, 26.80){\circle*{0.5}}
 \put( 22.90, 26.98){\circle*{0.4}}
 \put( 23.00, 26.96){\circle*{0.5}}
 \put( 23.00, 27.14){\circle*{0.4}}
 \put( 23.10, 27.12){\circle*{0.5}}
 \put( 23.10, 27.29){\circle*{0.4}}
 \put( 23.20, 27.28){\circle*{0.5}}
 \put( 23.20, 27.44){\circle*{0.4}}
 \put( 23.30, 27.44){\circle*{0.5}}
 \put( 23.30, 27.59){\circle*{0.4}}
 \put( 23.40, 27.59){\circle*{0.5}}
 \put( 23.40, 27.74){\circle*{0.4}}
 \put( 23.50, 27.75){\circle*{0.5}}
 \put( 23.50, 27.88){\circle*{0.4}}
 \put( 23.60, 27.91){\circle*{0.5}}
 \put( 23.60, 28.03){\circle*{0.4}}
 \put( 23.70, 28.06){\circle*{0.5}}
 \put( 23.70, 28.18){\circle*{0.4}}
 \put( 23.80, 28.21){\circle*{0.5}}
 \put( 23.80, 28.32){\circle*{0.4}}
 \put( 23.90, 28.37){\circle*{0.5}}
 \put( 23.90, 28.47){\circle*{0.4}}
 \put( 24.00, 28.52){\circle*{0.5}}
 \put( 24.00, 28.61){\circle*{0.4}}
 \put( 24.10, 28.67){\circle*{0.5}}
 \put( 24.10, 28.75){\circle*{0.4}}
 \put( 24.20, 28.82){\circle*{0.5}}
 \put( 24.20, 28.90){\circle*{0.4}}
 \put( 24.30, 28.97){\circle*{0.5}}
 \put( 24.30, 29.04){\circle*{0.4}}
 \put( 24.40, 29.12){\circle*{0.5}}
 \put( 24.40, 29.18){\circle*{0.4}}
 \put( 24.50, 29.27){\circle*{0.5}}
 \put( 24.50, 29.32){\circle*{0.4}}
 \put( 24.60, 29.42){\circle*{0.5}}
 \put( 24.60, 29.45){\circle*{0.4}}
 \put( 24.70, 29.56){\circle*{0.5}}
 \put( 24.70, 29.59){\circle*{0.4}}
 \put( 24.80, 29.71){\circle*{0.5}}
 \put( 24.80, 29.73){\circle*{0.4}}
 \put( 24.90, 29.86){\circle*{0.5}}
 \put( 24.90, 29.86){\circle*{0.4}}
 \put( 25.00, 30.00){\circle*{0.5}}
 \put( 25.00, 30.00){\circle*{0.4}}
 \put( 25.10, 30.14){\circle*{0.5}}
 \put( 25.10, 30.13){\circle*{0.4}}
 \put( 25.20, 30.29){\circle*{0.5}}
 \put( 25.20, 30.27){\circle*{0.4}}
 \put( 25.30, 30.43){\circle*{0.5}}
 \put( 25.30, 30.40){\circle*{0.4}}
 \put( 25.40, 30.57){\circle*{0.5}}
 \put( 25.40, 30.53){\circle*{0.4}}
 \put( 25.50, 30.71){\circle*{0.5}}
 \put( 25.50, 30.66){\circle*{0.4}}
 \put( 25.60, 30.85){\circle*{0.5}}
 \put( 25.60, 30.80){\circle*{0.4}}
 \put( 25.70, 30.99){\circle*{0.5}}
 \put( 25.70, 30.93){\circle*{0.4}}
 \put( 25.80, 31.13){\circle*{0.5}}
 \put( 25.80, 31.06){\circle*{0.4}}
 \put( 25.90, 31.27){\circle*{0.5}}
 \put( 25.90, 31.18){\circle*{0.4}}
 \put( 26.00, 31.41){\circle*{0.5}}
 \put( 26.00, 31.31){\circle*{0.4}}
 \put( 26.10, 31.54){\circle*{0.5}}
 \put( 26.10, 31.44){\circle*{0.4}}
 \put( 26.20, 31.68){\circle*{0.5}}
 \put( 26.20, 31.57){\circle*{0.4}}
 \put( 26.30, 31.82){\circle*{0.5}}
 \put( 26.30, 31.69){\circle*{0.4}}
 \put( 26.40, 31.95){\circle*{0.5}}
 \put( 26.40, 31.82){\circle*{0.4}}
 \put( 26.50, 32.09){\circle*{0.5}}
 \put( 26.50, 31.94){\circle*{0.4}}
 \put( 26.60, 32.22){\circle*{0.5}}
 \put( 26.60, 32.07){\circle*{0.4}}
 \put( 26.70, 32.35){\circle*{0.5}}
 \put( 26.70, 32.19){\circle*{0.4}}
 \put( 26.80, 32.49){\circle*{0.5}}
 \put( 26.80, 32.31){\circle*{0.4}}
 \put( 26.90, 32.62){\circle*{0.5}}
 \put( 26.90, 32.43){\circle*{0.4}}
 \put( 27.00, 32.75){\circle*{0.5}}
 \put( 27.00, 32.55){\circle*{0.4}}
 \put( 27.10, 32.88){\circle*{0.5}}
 \put( 27.10, 32.68){\circle*{0.4}}
 \put( 27.20, 33.01){\circle*{0.5}}
 \put( 27.20, 32.80){\circle*{0.4}}
 \put( 27.30, 33.14){\circle*{0.5}}
 \put( 27.30, 32.91){\circle*{0.4}}
 \put( 27.40, 33.27){\circle*{0.5}}
 \put( 27.40, 33.03){\circle*{0.4}}
 \put( 27.50, 33.40){\circle*{0.5}}
 \put( 27.50, 33.15){\circle*{0.4}}
 \put( 27.60, 33.53){\circle*{0.5}}
 \put( 27.60, 33.27){\circle*{0.4}}
 \put( 27.70, 33.65){\circle*{0.5}}
 \put( 27.70, 33.39){\circle*{0.4}}
 \put( 27.80, 33.78){\circle*{0.5}}
 \put( 27.80, 33.50){\circle*{0.4}}
 \put( 27.90, 33.91){\circle*{0.5}}
 \put( 27.90, 33.62){\circle*{0.4}}
 \put( 28.00, 34.03){\circle*{0.5}}
 \put( 28.00, 33.73){\circle*{0.4}}
 \put( 28.10, 34.16){\circle*{0.5}}
 \put( 28.10, 33.85){\circle*{0.4}}
 \put( 28.20, 34.28){\circle*{0.5}}
 \put( 28.20, 33.96){\circle*{0.4}}
 \put( 28.30, 34.41){\circle*{0.5}}
 \put( 28.30, 34.08){\circle*{0.4}}
 \put( 28.40, 34.53){\circle*{0.5}}
 \put( 28.40, 34.19){\circle*{0.4}}
 \put( 28.50, 34.65){\circle*{0.5}}
 \put( 28.50, 34.30){\circle*{0.4}}
 \put( 28.60, 34.78){\circle*{0.5}}
 \put( 28.60, 34.42){\circle*{0.4}}
 \put( 28.70, 34.90){\circle*{0.5}}
 \put( 28.70, 34.53){\circle*{0.4}}
 \put( 28.80, 35.02){\circle*{0.5}}
 \put( 28.80, 34.64){\circle*{0.4}}
 \put( 28.90, 35.14){\circle*{0.5}}
 \put( 28.90, 34.75){\circle*{0.4}}
 \put( 29.00, 35.26){\circle*{0.5}}
 \put( 29.00, 34.86){\circle*{0.4}}
 \put( 29.10, 35.38){\circle*{0.5}}
 \put( 29.10, 34.97){\circle*{0.4}}
 \put( 29.20, 35.50){\circle*{0.5}}
 \put( 29.20, 35.08){\circle*{0.4}}
 \put( 29.30, 35.62){\circle*{0.5}}
 \put( 29.30, 35.19){\circle*{0.4}}
 \put( 29.40, 35.74){\circle*{0.5}}
 \put( 29.40, 35.29){\circle*{0.4}}
 \put( 29.50, 35.86){\circle*{0.5}}
 \put( 29.50, 35.40){\circle*{0.4}}
 \put( 29.60, 35.97){\circle*{0.5}}
 \put( 29.60, 35.51){\circle*{0.4}}
 \put( 29.70, 36.09){\circle*{0.5}}
 \put( 29.70, 35.62){\circle*{0.4}}
 \put( 29.80, 36.21){\circle*{0.5}}
 \put( 29.80, 35.72){\circle*{0.4}}
 \put( 29.90, 36.32){\circle*{0.5}}
 \put( 29.90, 35.83){\circle*{0.4}}
 \put( 30.00, 36.44){\circle*{0.5}}
 \put( 30.00, 35.93){\circle*{0.4}}
 \put( 30.10, 36.55){\circle*{0.5}}
 \put( 30.10, 36.04){\circle*{0.4}}
 \put( 30.20, 36.67){\circle*{0.5}}
 \put( 30.20, 36.14){\circle*{0.4}}
 \put( 30.30, 36.78){\circle*{0.5}}
 \put( 30.30, 36.24){\circle*{0.4}}
 \put( 30.40, 36.90){\circle*{0.5}}
 \put( 30.40, 36.35){\circle*{0.4}}
 \put( 30.50, 37.01){\circle*{0.5}}
 \put( 30.50, 36.45){\circle*{0.4}}
 \put( 30.60, 37.12){\circle*{0.5}}
 \put( 30.60, 36.55){\circle*{0.4}}
 \put( 30.70, 37.24){\circle*{0.5}}
 \put( 30.70, 36.66){\circle*{0.4}}
 \put( 30.80, 37.35){\circle*{0.5}}
 \put( 30.80, 36.76){\circle*{0.4}}
 \put( 30.90, 37.46){\circle*{0.5}}
 \put( 30.90, 36.86){\circle*{0.4}}
 \put( 31.00, 37.57){\circle*{0.5}}
 \put( 31.00, 36.96){\circle*{0.4}}
 \put( 31.10, 37.68){\circle*{0.5}}
 \put( 31.10, 37.06){\circle*{0.4}}
 \put( 31.20, 37.79){\circle*{0.5}}
 \put( 31.20, 37.16){\circle*{0.4}}
 \put( 31.30, 37.90){\circle*{0.5}}
 \put( 31.30, 37.26){\circle*{0.4}}
 \put( 31.40, 38.01){\circle*{0.5}}
 \put( 31.40, 37.36){\circle*{0.4}}
 \put( 31.50, 38.12){\circle*{0.5}}
 \put( 31.50, 37.46){\circle*{0.4}}
 \put( 31.60, 38.23){\circle*{0.5}}
 \put( 31.60, 37.55){\circle*{0.4}}
 \put( 31.70, 38.34){\circle*{0.5}}
 \put( 31.70, 37.65){\circle*{0.4}}
 \put( 31.80, 38.44){\circle*{0.5}}
 \put( 31.80, 37.75){\circle*{0.4}}
 \put( 31.90, 38.55){\circle*{0.5}}
 \put( 31.90, 37.85){\circle*{0.4}}
 \put( 32.00, 38.66){\circle*{0.5}}
 \put( 32.00, 37.94){\circle*{0.4}}
 \put( 32.10, 38.77){\circle*{0.5}}
 \put( 32.10, 38.04){\circle*{0.4}}
 \put( 32.20, 38.87){\circle*{0.5}}
 \put( 32.20, 38.13){\circle*{0.4}}
 \put( 32.30, 38.98){\circle*{0.5}}
 \put( 32.30, 38.23){\circle*{0.4}}
 \put( 32.40, 39.08){\circle*{0.5}}
 \put( 32.40, 38.32){\circle*{0.4}}
 \put( 32.50, 39.19){\circle*{0.5}}
 \put( 32.50, 38.42){\circle*{0.4}}
 \put( 32.60, 39.29){\circle*{0.5}}
 \put( 32.60, 38.51){\circle*{0.4}}
 \put( 32.70, 39.40){\circle*{0.5}}
 \put( 32.70, 38.61){\circle*{0.4}}
 \put( 32.80, 39.50){\circle*{0.5}}
 \put( 32.80, 38.70){\circle*{0.4}}
 \put( 32.90, 39.61){\circle*{0.5}}
 \put( 32.90, 38.79){\circle*{0.4}}
 \put( 33.00, 39.71){\circle*{0.5}}
 \put( 33.00, 38.89){\circle*{0.4}}
 \put( 33.10, 39.81){\circle*{0.5}}
 \put( 33.10, 38.98){\circle*{0.4}}
 \put( 33.20, 39.91){\circle*{0.5}}
 \put( 33.20, 39.07){\circle*{0.4}}
 \put( 33.30, 40.02){\circle*{0.5}}
 \put( 33.30, 39.16){\circle*{0.4}}
 \put( 33.40, 40.12){\circle*{0.5}}
 \put( 33.40, 39.25){\circle*{0.4}}
 \put( 33.50, 40.22){\circle*{0.5}}
 \put( 33.50, 39.34){\circle*{0.4}}
 \put( 33.60, 40.32){\circle*{0.5}}
 \put( 33.60, 39.44){\circle*{0.4}}
 \put( 33.70, 40.42){\circle*{0.5}}
 \put( 33.70, 39.53){\circle*{0.4}}
 \put( 33.80, 40.52){\circle*{0.5}}
 \put( 33.80, 39.62){\circle*{0.4}}
 \put( 33.90, 40.62){\circle*{0.5}}
 \put( 33.90, 39.71){\circle*{0.4}}
 \put( 34.00, 40.72){\circle*{0.5}}
 \put( 34.00, 39.79){\circle*{0.4}}
 \put( 34.10, 40.82){\circle*{0.5}}
 \put( 34.10, 39.88){\circle*{0.4}}
 \put( 34.20, 40.92){\circle*{0.5}}
 \put( 34.20, 39.97){\circle*{0.4}}
 \put( 34.30, 41.02){\circle*{0.5}}
 \put( 34.30, 40.06){\circle*{0.4}}
 \put( 34.40, 41.12){\circle*{0.5}}
 \put( 34.40, 40.15){\circle*{0.4}}
 \put( 34.50, 41.21){\circle*{0.5}}
 \put( 34.50, 40.24){\circle*{0.4}}
 \put( 34.60, 41.31){\circle*{0.5}}
 \put( 34.60, 40.32){\circle*{0.4}}
 \put( 34.70, 41.41){\circle*{0.5}}
 \put( 34.70, 40.41){\circle*{0.4}}
 \put( 34.80, 41.51){\circle*{0.5}}
 \put( 34.80, 40.50){\circle*{0.4}}
 \put( 34.90, 41.60){\circle*{0.5}}
 \put( 34.90, 40.58){\circle*{0.4}}
 \put( 35.00, 41.70){\circle*{0.5}}
 \put( 35.00, 40.67){\circle*{0.4}}
 \put( 35.10, 41.80){\circle*{0.5}}
 \put( 35.10, 40.75){\circle*{0.4}}
 \put( 35.20, 41.89){\circle*{0.5}}
 \put( 35.20, 40.84){\circle*{0.4}}
 \put( 35.30, 41.99){\circle*{0.5}}
 \put( 35.30, 40.92){\circle*{0.4}}
 \put( 35.40, 42.08){\circle*{0.5}}
 \put( 35.40, 41.01){\circle*{0.4}}
 \put( 35.50, 42.18){\circle*{0.5}}
 \put( 35.50, 41.09){\circle*{0.4}}
 \put( 35.60, 42.27){\circle*{0.5}}
 \put( 35.60, 41.18){\circle*{0.4}}
 \put( 35.70, 42.36){\circle*{0.5}}
 \put( 35.70, 41.26){\circle*{0.4}}
 \put( 35.80, 42.46){\circle*{0.5}}
 \put( 35.80, 41.35){\circle*{0.4}}
 \put( 35.90, 42.55){\circle*{0.5}}
 \put( 35.90, 41.43){\circle*{0.4}}
 \put( 36.00, 42.65){\circle*{0.5}}
 \put( 36.00, 41.51){\circle*{0.4}}
 \put( 36.10, 42.74){\circle*{0.5}}
 \put( 36.10, 41.59){\circle*{0.4}}
 \put( 36.20, 42.83){\circle*{0.5}}
 \put( 36.20, 41.68){\circle*{0.4}}
 \put( 36.30, 42.92){\circle*{0.5}}
 \put( 36.30, 41.76){\circle*{0.4}}
 \put( 36.40, 43.02){\circle*{0.5}}
 \put( 36.40, 41.84){\circle*{0.4}}
 \put( 36.50, 43.11){\circle*{0.5}}
 \put( 36.50, 41.92){\circle*{0.4}}
 \put( 36.60, 43.20){\circle*{0.5}}
 \put( 36.60, 42.00){\circle*{0.4}}
 \put( 36.70, 43.29){\circle*{0.5}}
 \put( 36.70, 42.08){\circle*{0.4}}
 \put( 36.80, 43.38){\circle*{0.5}}
 \put( 36.80, 42.17){\circle*{0.4}}
 \put( 36.90, 43.47){\circle*{0.5}}
 \put( 36.90, 42.25){\circle*{0.4}}
 \put( 37.00, 43.56){\circle*{0.5}}
 \put( 37.00, 42.33){\circle*{0.4}}
 \put( 37.10, 43.65){\circle*{0.5}}
 \put( 37.10, 42.41){\circle*{0.4}}
 \put( 37.20, 43.74){\circle*{0.5}}
 \put( 37.20, 42.49){\circle*{0.4}}
 \put( 37.30, 43.83){\circle*{0.5}}
 \put( 37.30, 42.56){\circle*{0.4}}
 \put( 37.40, 43.92){\circle*{0.5}}
 \put( 37.40, 42.64){\circle*{0.4}}
 \put( 37.50, 44.01){\circle*{0.5}}
 \put( 37.50, 42.72){\circle*{0.4}}
 \put( 37.60, 44.10){\circle*{0.5}}
 \put( 37.60, 42.80){\circle*{0.4}}
 \put( 37.70, 44.19){\circle*{0.5}}
 \put( 37.70, 42.88){\circle*{0.4}}
 \put( 37.80, 44.27){\circle*{0.5}}
 \put( 37.80, 42.96){\circle*{0.4}}
 \put( 37.90, 44.36){\circle*{0.5}}
 \put( 37.90, 43.04){\circle*{0.4}}
 \put( 38.00, 44.45){\circle*{0.5}}
 \put( 38.00, 43.11){\circle*{0.4}}
 \put( 38.10, 44.54){\circle*{0.5}}
 \put( 38.10, 43.19){\circle*{0.4}}
 \put( 38.20, 44.62){\circle*{0.5}}
 \put( 38.20, 43.27){\circle*{0.4}}
 \put( 38.30, 44.71){\circle*{0.5}}
 \put( 38.30, 43.34){\circle*{0.4}}
 \put( 38.40, 44.80){\circle*{0.5}}
 \put( 38.40, 43.42){\circle*{0.4}}
 \put( 38.50, 44.88){\circle*{0.5}}
 \put( 38.50, 43.50){\circle*{0.4}}
 \put( 38.60, 44.97){\circle*{0.5}}
 \put( 38.60, 43.57){\circle*{0.4}}
 \put( 38.70, 45.05){\circle*{0.5}}
 \put( 38.70, 43.65){\circle*{0.4}}
 \put( 38.80, 45.14){\circle*{0.5}}
 \put( 38.80, 43.72){\circle*{0.4}}
 \put( 38.90, 45.23){\circle*{0.5}}
 \put( 38.90, 43.80){\circle*{0.4}}
 \put( 39.00, 45.31){\circle*{0.5}}
 \put( 39.00, 43.87){\circle*{0.4}}
 \put( 39.10, 45.40){\circle*{0.5}}
 \put( 39.10, 43.95){\circle*{0.4}}
 \put( 39.20, 45.48){\circle*{0.5}}
 \put( 39.20, 44.02){\circle*{0.4}}
 \put( 39.30, 45.56){\circle*{0.5}}
 \put( 39.30, 44.10){\circle*{0.4}}
 \put( 39.40, 45.65){\circle*{0.5}}
 \put( 39.40, 44.17){\circle*{0.4}}
 \put( 39.50, 45.73){\circle*{0.5}}
 \put( 39.50, 44.25){\circle*{0.4}}
 \put( 39.60, 45.82){\circle*{0.5}}
 \put( 39.60, 44.32){\circle*{0.4}}
 \put( 39.70, 45.90){\circle*{0.5}}
 \put( 39.70, 44.39){\circle*{0.4}}
 \put( 39.80, 45.98){\circle*{0.5}}
 \put( 39.80, 44.47){\circle*{0.4}}
 \put( 39.90, 46.06){\circle*{0.5}}
 \put( 39.90, 44.54){\circle*{0.4}}
 \put( 40.00, 46.15){\circle*{0.5}}
 \put( 40.00, 44.61){\circle*{0.4}}
 \put( 40.10, 46.23){\circle*{0.5}}
 \put( 40.10, 44.68){\circle*{0.4}}
 \put( 40.20, 46.31){\circle*{0.5}}
 \put( 40.20, 44.76){\circle*{0.4}}
 \put( 40.30, 46.39){\circle*{0.5}}
 \put( 40.30, 44.83){\circle*{0.4}}
 \put( 40.40, 46.47){\circle*{0.5}}
 \put( 40.40, 44.90){\circle*{0.4}}
 \put( 40.50, 46.56){\circle*{0.5}}
 \put( 40.50, 44.97){\circle*{0.4}}
 \put( 40.60, 46.64){\circle*{0.5}}
 \put( 40.60, 45.04){\circle*{0.4}}
 \put( 40.70, 46.72){\circle*{0.5}}
 \put( 40.70, 45.12){\circle*{0.4}}
 \put( 40.80, 46.80){\circle*{0.5}}
 \put( 40.80, 45.19){\circle*{0.4}}
 \put( 40.90, 46.88){\circle*{0.5}}
 \put( 40.90, 45.26){\circle*{0.4}}
 \put( 41.00, 46.96){\circle*{0.5}}
 \put( 41.00, 45.33){\circle*{0.4}}
 \put( 41.10, 47.04){\circle*{0.5}}
 \put( 41.10, 45.40){\circle*{0.4}}
 \put( 41.20, 47.12){\circle*{0.5}}
 \put( 41.20, 45.47){\circle*{0.4}}
 \put( 41.30, 47.20){\circle*{0.5}}
 \put( 41.30, 45.54){\circle*{0.4}}
 \put( 41.40, 47.28){\circle*{0.5}}
 \put( 41.40, 45.61){\circle*{0.4}}
 \put( 41.50, 47.36){\circle*{0.5}}
 \put( 41.50, 45.68){\circle*{0.4}}
 \put( 41.60, 47.44){\circle*{0.5}}
 \put( 41.60, 45.75){\circle*{0.4}}
 \put( 41.70, 47.52){\circle*{0.5}}
 \put( 41.70, 45.82){\circle*{0.4}}
 \put( 41.80, 47.59){\circle*{0.5}}
 \put( 41.80, 45.88){\circle*{0.4}}
 \put( 41.90, 47.67){\circle*{0.5}}
 \put( 41.90, 45.95){\circle*{0.4}}
 \put( 42.00, 47.75){\circle*{0.5}}
 \put( 42.00, 46.02){\circle*{0.4}}
 \put( 42.10, 47.83){\circle*{0.5}}
 \put( 42.10, 46.09){\circle*{0.4}}
 \put( 42.20, 47.91){\circle*{0.5}}
 \put( 42.20, 46.16){\circle*{0.4}}
 \put( 42.30, 47.98){\circle*{0.5}}
 \put( 42.30, 46.23){\circle*{0.4}}
 \put( 42.40, 48.06){\circle*{0.5}}
 \put( 42.40, 46.29){\circle*{0.4}}
 \put( 42.50, 48.14){\circle*{0.5}}
 \put( 42.50, 46.36){\circle*{0.4}}
 \put( 42.60, 48.21){\circle*{0.5}}
 \put( 42.60, 46.43){\circle*{0.4}}
 \put( 42.70, 48.29){\circle*{0.5}}
 \put( 42.70, 46.50){\circle*{0.4}}
 \put( 42.80, 48.37){\circle*{0.5}}
 \put( 42.80, 46.56){\circle*{0.4}}
 \put( 42.90, 48.44){\circle*{0.5}}
 \put( 42.90, 46.63){\circle*{0.4}}
 \put( 43.00, 48.52){\circle*{0.5}}
 \put( 43.00, 46.70){\circle*{0.4}}
 \put( 43.10, 48.60){\circle*{0.5}}
 \put( 43.10, 46.76){\circle*{0.4}}
 \put( 43.20, 48.67){\circle*{0.5}}
 \put( 43.20, 46.83){\circle*{0.4}}
 \put( 43.30, 48.75){\circle*{0.5}}
 \put( 43.30, 46.89){\circle*{0.4}}
 \put( 43.40, 48.82){\circle*{0.5}}
 \put( 43.40, 46.96){\circle*{0.4}}
 \put( 43.50, 48.90){\circle*{0.5}}
 \put( 43.50, 47.03){\circle*{0.4}}
 \put( 43.60, 48.97){\circle*{0.5}}
 \put( 43.60, 47.09){\circle*{0.4}}
 \put( 43.70, 49.05){\circle*{0.5}}
 \put( 43.70, 47.16){\circle*{0.4}}
 \put( 43.80, 49.12){\circle*{0.5}}
 \put( 43.80, 47.22){\circle*{0.4}}
 \put( 43.90, 49.20){\circle*{0.5}}
 \put( 43.90, 47.29){\circle*{0.4}}
 \put( 44.00, 49.27){\circle*{0.5}}
 \put( 44.00, 47.35){\circle*{0.4}}
 \put( 44.10, 49.34){\circle*{0.5}}
 \put( 44.10, 47.42){\circle*{0.4}}
 \put( 44.20, 49.42){\circle*{0.5}}
 \put( 44.20, 47.48){\circle*{0.4}}
 \put( 44.30, 49.49){\circle*{0.5}}
 \put( 44.30, 47.55){\circle*{0.4}}
 \put( 44.40, 49.56){\circle*{0.5}}
 \put( 44.40, 47.61){\circle*{0.4}}
 \put( 44.50, 49.64){\circle*{0.5}}
 \put( 44.50, 47.67){\circle*{0.4}}
 \put( 44.60, 49.71){\circle*{0.5}}
 \put( 44.60, 47.74){\circle*{0.4}}
 \put( 44.70, 49.78){\circle*{0.5}}
 \put( 44.70, 47.80){\circle*{0.4}}
 \put( 44.80, 49.86){\circle*{0.5}}
 \put( 44.80, 47.86){\circle*{0.4}}
 \put( 44.90, 49.93){\circle*{0.5}}
 \put( 44.90, 47.93){\circle*{0.4}}
 \put( 45.00, 50.00){\circle*{0.5}}
 \put( 45.00, 47.99){\circle*{0.4}}
 \put( 45.10, 50.07){\circle*{0.5}}
 \put( 45.10, 48.05){\circle*{0.4}}
 \put( 45.20, 50.14){\circle*{0.5}}
 \put( 45.20, 48.12){\circle*{0.4}}
 \put( 45.30, 50.22){\circle*{0.5}}
 \put( 45.30, 48.18){\circle*{0.4}}
 \put( 45.40, 50.29){\circle*{0.5}}
 \put( 45.40, 48.24){\circle*{0.4}}
 \put( 45.50, 50.36){\circle*{0.5}}
 \put( 45.50, 48.30){\circle*{0.4}}
 \put( 45.60, 50.43){\circle*{0.5}}
 \put( 45.60, 48.36){\circle*{0.4}}
 \put( 45.70, 50.50){\circle*{0.5}}
 \put( 45.70, 48.43){\circle*{0.4}}
 \put( 45.80, 50.57){\circle*{0.5}}
 \put( 45.80, 48.49){\circle*{0.4}}
 \put( 45.90, 50.64){\circle*{0.5}}
 \put( 45.90, 48.55){\circle*{0.4}}
 \put( 46.00, 50.71){\circle*{0.5}}
 \put( 46.00, 48.61){\circle*{0.4}}
 \put( 46.10, 50.78){\circle*{0.5}}
 \put( 46.10, 48.67){\circle*{0.4}}
 \put( 46.20, 50.85){\circle*{0.5}}
 \put( 46.20, 48.73){\circle*{0.4}}
 \put( 46.30, 50.92){\circle*{0.5}}
 \put( 46.30, 48.79){\circle*{0.4}}
 \put( 46.40, 50.99){\circle*{0.5}}
 \put( 46.40, 48.86){\circle*{0.4}}
 \put( 46.50, 51.06){\circle*{0.5}}
 \put( 46.50, 48.92){\circle*{0.4}}
 \put( 46.60, 51.13){\circle*{0.5}}
 \put( 46.60, 48.98){\circle*{0.4}}
 \put( 46.70, 51.20){\circle*{0.5}}
 \put( 46.70, 49.04){\circle*{0.4}}
 \put( 46.80, 51.27){\circle*{0.5}}
 \put( 46.80, 49.10){\circle*{0.4}}
 \put( 46.90, 51.34){\circle*{0.5}}
 \put( 46.90, 49.16){\circle*{0.4}}
 \put( 47.00, 51.41){\circle*{0.5}}
 \put( 47.00, 49.22){\circle*{0.4}}
 \put( 47.10, 51.48){\circle*{0.5}}
 \put( 47.10, 49.28){\circle*{0.4}}
 \put( 47.20, 51.54){\circle*{0.5}}
 \put( 47.20, 49.34){\circle*{0.4}}
 \put( 47.30, 51.61){\circle*{0.5}}
 \put( 47.30, 49.39){\circle*{0.4}}
 \put( 47.40, 51.68){\circle*{0.5}}
 \put( 47.40, 49.45){\circle*{0.4}}
 \put( 47.50, 51.75){\circle*{0.5}}
 \put( 47.50, 49.51){\circle*{0.4}}
 \put( 47.60, 51.82){\circle*{0.5}}
 \put( 47.60, 49.57){\circle*{0.4}}
 \put( 47.70, 51.88){\circle*{0.5}}
 \put( 47.70, 49.63){\circle*{0.4}}
 \put( 47.80, 51.95){\circle*{0.5}}
 \put( 47.80, 49.69){\circle*{0.4}}
 \put( 47.90, 52.02){\circle*{0.5}}
 \put( 47.90, 49.75){\circle*{0.4}}
 \put( 48.00, 52.09){\circle*{0.5}}
 \put( 48.00, 49.81){\circle*{0.4}}
 \put( 48.10, 52.15){\circle*{0.5}}
 \put( 48.10, 49.86){\circle*{0.4}}
 \put( 48.20, 52.22){\circle*{0.5}}
 \put( 48.20, 49.92){\circle*{0.4}}
 \put( 48.30, 52.29){\circle*{0.5}}
 \put( 48.30, 49.98){\circle*{0.4}}
 \put( 48.40, 52.35){\circle*{0.5}}
 \put( 48.40, 50.04){\circle*{0.4}}
 \put( 48.50, 52.42){\circle*{0.5}}
 \put( 48.50, 50.10){\circle*{0.4}}
 \put( 48.60, 52.49){\circle*{0.5}}
 \put( 48.60, 50.15){\circle*{0.4}}
 \put( 48.70, 52.55){\circle*{0.5}}
 \put( 48.70, 50.21){\circle*{0.4}}
 \put( 48.80, 52.62){\circle*{0.5}}
 \put( 48.80, 50.27){\circle*{0.4}}
 \put( 48.90, 52.68){\circle*{0.5}}
 \put( 48.90, 50.33){\circle*{0.4}}
 \put( 49.00, 52.75){\circle*{0.5}}
 \put( 49.00, 50.38){\circle*{0.4}}
 \put( 49.10, 52.82){\circle*{0.5}}
 \put( 49.10, 50.44){\circle*{0.4}}
 \put( 49.20, 52.88){\circle*{0.5}}
 \put( 49.20, 50.50){\circle*{0.4}}
 \put( 49.30, 52.95){\circle*{0.5}}
 \put( 49.30, 50.55){\circle*{0.4}}
 \put( 49.40, 53.01){\circle*{0.5}}
 \put( 49.40, 50.61){\circle*{0.4}}
 \put( 49.50, 53.08){\circle*{0.5}}
 \put( 49.50, 50.66){\circle*{0.4}}
 \put( 49.60, 53.14){\circle*{0.5}}
 \put( 49.60, 50.72){\circle*{0.4}}
 \put( 49.70, 53.21){\circle*{0.5}}
 \put( 49.70, 50.78){\circle*{0.4}}
 \put( 49.80, 53.27){\circle*{0.5}}
 \put( 49.80, 50.83){\circle*{0.4}}
 \put( 49.90, 53.33){\circle*{0.5}}
 \put( 49.90, 50.89){\circle*{0.4}}
 \put( 50.00, 53.40){\circle*{0.5}}
 \put( 50.00, 50.94){\circle*{0.4}}
 \put( 50.10, 53.46){\circle*{0.5}}
 \put( 50.10, 51.00){\circle*{0.4}}
 \put( 50.20, 53.53){\circle*{0.5}}
 \put( 50.20, 51.06){\circle*{0.4}}
 \put( 50.30, 53.59){\circle*{0.5}}
 \put( 50.30, 51.11){\circle*{0.4}}
 \put( 50.40, 53.65){\circle*{0.5}}
 \put( 50.40, 51.17){\circle*{0.4}}
 \put( 50.50, 53.72){\circle*{0.5}}
 \put( 50.50, 51.22){\circle*{0.4}}
 \put( 50.60, 53.78){\circle*{0.5}}
 \put( 50.60, 51.28){\circle*{0.4}}
 \put( 50.70, 53.84){\circle*{0.5}}
 \put( 50.70, 51.33){\circle*{0.4}}
 \put( 50.80, 53.91){\circle*{0.5}}
 \put( 50.80, 51.38){\circle*{0.4}}
 \put( 50.90, 53.97){\circle*{0.5}}
 \put( 50.90, 51.44){\circle*{0.4}}
 \put( 51.00, 54.03){\circle*{0.5}}
 \put( 51.00, 51.49){\circle*{0.4}}
 \put( 51.10, 54.10){\circle*{0.5}}
 \put( 51.10, 51.55){\circle*{0.4}}
 \put( 51.20, 54.16){\circle*{0.5}}
 \put( 51.20, 51.60){\circle*{0.4}}
 \put( 51.30, 54.22){\circle*{0.5}}
 \put( 51.30, 51.66){\circle*{0.4}}
 \put( 51.40, 54.28){\circle*{0.5}}
 \put( 51.40, 51.71){\circle*{0.4}}
 \put( 51.50, 54.34){\circle*{0.5}}
 \put( 51.50, 51.76){\circle*{0.4}}
 \put( 51.60, 54.41){\circle*{0.5}}
 \put( 51.60, 51.82){\circle*{0.4}}
 \put( 51.70, 54.47){\circle*{0.5}}
 \put( 51.70, 51.87){\circle*{0.4}}
 \put( 51.80, 54.53){\circle*{0.5}}
 \put( 51.80, 51.92){\circle*{0.4}}
 \put( 51.90, 54.59){\circle*{0.5}}
 \put( 51.90, 51.98){\circle*{0.4}}
 \put( 52.00, 54.65){\circle*{0.5}}
 \put( 52.00, 52.03){\circle*{0.4}}
 \put( 52.10, 54.71){\circle*{0.5}}
 \put( 52.10, 52.08){\circle*{0.4}}
 \put( 52.20, 54.78){\circle*{0.5}}
 \put( 52.20, 52.14){\circle*{0.4}}
 \put( 52.30, 54.84){\circle*{0.5}}
 \put( 52.30, 52.19){\circle*{0.4}}
 \put( 52.40, 54.90){\circle*{0.5}}
 \put( 52.40, 52.24){\circle*{0.4}}
 \put( 52.50, 54.96){\circle*{0.5}}
 \put( 52.50, 52.29){\circle*{0.4}}
 \put( 52.60, 55.02){\circle*{0.5}}
 \put( 52.60, 52.35){\circle*{0.4}}
 \put( 52.70, 55.08){\circle*{0.5}}
 \put( 52.70, 52.40){\circle*{0.4}}
 \put( 52.80, 55.14){\circle*{0.5}}
 \put( 52.80, 52.45){\circle*{0.4}}
 \put( 52.90, 55.20){\circle*{0.5}}
 \put( 52.90, 52.50){\circle*{0.4}}
 \put( 53.00, 55.26){\circle*{0.5}}
 \put( 53.00, 52.55){\circle*{0.4}}
 \put( 53.10, 55.32){\circle*{0.5}}
 \put( 53.10, 52.61){\circle*{0.4}}
 \put( 53.20, 55.38){\circle*{0.5}}
 \put( 53.20, 52.66){\circle*{0.4}}
 \put( 53.30, 55.44){\circle*{0.5}}
 \put( 53.30, 52.71){\circle*{0.4}}
 \put( 53.40, 55.50){\circle*{0.5}}
 \put( 53.40, 52.76){\circle*{0.4}}
 \put( 53.50, 55.56){\circle*{0.5}}
 \put( 53.50, 52.81){\circle*{0.4}}
 \put( 53.60, 55.62){\circle*{0.5}}
 \put( 53.60, 52.86){\circle*{0.4}}
 \put( 53.70, 55.68){\circle*{0.5}}
 \put( 53.70, 52.91){\circle*{0.4}}
 \put( 53.80, 55.74){\circle*{0.5}}
 \put( 53.80, 52.97){\circle*{0.4}}
 \put( 53.90, 55.80){\circle*{0.5}}
 \put( 53.90, 53.02){\circle*{0.4}}
 \put( 54.00, 55.86){\circle*{0.5}}
 \put( 54.00, 53.07){\circle*{0.4}}
 \put( 54.10, 55.91){\circle*{0.5}}
 \put( 54.10, 53.12){\circle*{0.4}}
 \put( 54.20, 55.97){\circle*{0.5}}
 \put( 54.20, 53.17){\circle*{0.4}}
 \put( 54.30, 56.03){\circle*{0.5}}
 \put( 54.30, 53.22){\circle*{0.4}}
 \put( 54.40, 56.09){\circle*{0.5}}
 \put( 54.40, 53.27){\circle*{0.4}}
 \put( 54.50, 56.15){\circle*{0.5}}
 \put( 54.50, 53.32){\circle*{0.4}}
 \put( 54.60, 56.21){\circle*{0.5}}
 \put( 54.60, 53.37){\circle*{0.4}}
 \put( 54.70, 56.26){\circle*{0.5}}
 \put( 54.70, 53.42){\circle*{0.4}}
 \put( 54.80, 56.32){\circle*{0.5}}
 \put( 54.80, 53.47){\circle*{0.4}}
 \put( 54.90, 56.38){\circle*{0.5}}
 \put( 54.90, 53.52){\circle*{0.4}}
 \put( 55.00, 56.44){\circle*{0.5}}
 \put( 55.00, 53.57){\circle*{0.4}}
 \put( 55.10, 56.50){\circle*{0.5}}
 \put( 55.10, 53.62){\circle*{0.4}}
 \put( 55.20, 56.55){\circle*{0.5}}
 \put( 55.20, 53.67){\circle*{0.4}}
 \put( 55.30, 56.61){\circle*{0.5}}
 \put( 55.30, 53.72){\circle*{0.4}}
 \put( 55.40, 56.67){\circle*{0.5}}
 \put( 55.40, 53.77){\circle*{0.4}}
 \put( 55.50, 56.73){\circle*{0.5}}
 \put( 55.50, 53.82){\circle*{0.4}}
 \put( 55.60, 56.78){\circle*{0.5}}
 \put( 55.60, 53.87){\circle*{0.4}}
 \put( 55.70, 56.84){\circle*{0.5}}
 \put( 55.70, 53.92){\circle*{0.4}}
 \put( 55.80, 56.90){\circle*{0.5}}
 \put( 55.80, 53.96){\circle*{0.4}}
 \put( 55.90, 56.95){\circle*{0.5}}
 \put( 55.90, 54.01){\circle*{0.4}}
 \put( 56.00, 57.01){\circle*{0.5}}
 \put( 56.00, 54.06){\circle*{0.4}}
 \put( 56.10, 57.07){\circle*{0.5}}
 \put( 56.10, 54.11){\circle*{0.4}}
 \put( 56.20, 57.12){\circle*{0.5}}
 \put( 56.20, 54.16){\circle*{0.4}}
 \put( 56.30, 57.18){\circle*{0.5}}
 \put( 56.30, 54.21){\circle*{0.4}}
 \put( 56.40, 57.24){\circle*{0.5}}
 \put( 56.40, 54.26){\circle*{0.4}}
 \put( 56.50, 57.29){\circle*{0.5}}
 \put( 56.50, 54.30){\circle*{0.4}}
 \put( 56.60, 57.35){\circle*{0.5}}
 \put( 56.60, 54.35){\circle*{0.4}}
 \put( 56.70, 57.40){\circle*{0.5}}
 \put( 56.70, 54.40){\circle*{0.4}}
 \put( 56.80, 57.46){\circle*{0.5}}
 \put( 56.80, 54.45){\circle*{0.4}}
 \put( 56.90, 57.51){\circle*{0.5}}
 \put( 56.90, 54.50){\circle*{0.4}}
 \put( 57.00, 57.57){\circle*{0.5}}
 \put( 57.00, 54.54){\circle*{0.4}}
 \put( 57.10, 57.63){\circle*{0.5}}
 \put( 57.10, 54.59){\circle*{0.4}}
 \put( 57.20, 57.68){\circle*{0.5}}
 \put( 57.20, 54.64){\circle*{0.4}}
 \put( 57.30, 57.74){\circle*{0.5}}
 \put( 57.30, 54.69){\circle*{0.4}}
 \put( 57.40, 57.79){\circle*{0.5}}
 \put( 57.40, 54.73){\circle*{0.4}}
 \put( 57.50, 57.85){\circle*{0.5}}
 \put( 57.50, 54.78){\circle*{0.4}}
 \put( 57.60, 57.90){\circle*{0.5}}
 \put( 57.60, 54.83){\circle*{0.4}}
 \put( 57.70, 57.96){\circle*{0.5}}
 \put( 57.70, 54.87){\circle*{0.4}}
 \put( 57.80, 58.01){\circle*{0.5}}
 \put( 57.80, 54.92){\circle*{0.4}}
 \put( 57.90, 58.07){\circle*{0.5}}
 \put( 57.90, 54.97){\circle*{0.4}}
 \put( 58.00, 58.12){\circle*{0.5}}
 \put( 58.00, 55.02){\circle*{0.4}}
 \put( 58.10, 58.17){\circle*{0.5}}
 \put( 58.10, 55.06){\circle*{0.4}}
 \put( 58.20, 58.23){\circle*{0.5}}
 \put( 58.20, 55.11){\circle*{0.4}}
 \put( 58.30, 58.28){\circle*{0.5}}
 \put( 58.30, 55.16){\circle*{0.4}}
 \put( 58.40, 58.34){\circle*{0.5}}
 \put( 58.40, 55.20){\circle*{0.4}}
 \put( 58.50, 58.39){\circle*{0.5}}
 \put( 58.50, 55.25){\circle*{0.4}}
 \put( 58.60, 58.44){\circle*{0.5}}
 \put( 58.60, 55.29){\circle*{0.4}}
 \put( 58.70, 58.50){\circle*{0.5}}
 \put( 58.70, 55.34){\circle*{0.4}}
 \put( 58.80, 58.55){\circle*{0.5}}
 \put( 58.80, 55.39){\circle*{0.4}}
 \put( 58.90, 58.61){\circle*{0.5}}
 \put( 58.90, 55.43){\circle*{0.4}}
 \put( 59.00, 58.66){\circle*{0.5}}
 \put( 59.00, 55.48){\circle*{0.4}}
 \put( 59.10, 58.71){\circle*{0.5}}
 \put( 59.10, 55.52){\circle*{0.4}}
 \put( 59.20, 58.77){\circle*{0.5}}
 \put( 59.20, 55.57){\circle*{0.4}}
 \put( 59.30, 58.82){\circle*{0.5}}
 \put( 59.30, 55.62){\circle*{0.4}}
 \put( 59.40, 58.87){\circle*{0.5}}
 \put( 59.40, 55.66){\circle*{0.4}}
 \put( 59.50, 58.93){\circle*{0.5}}
 \put( 59.50, 55.71){\circle*{0.4}}
 \put( 59.60, 58.98){\circle*{0.5}}
 \put( 59.60, 55.75){\circle*{0.4}}
 \put( 59.70, 59.03){\circle*{0.5}}
 \put( 59.70, 55.80){\circle*{0.4}}
 \put( 59.80, 59.08){\circle*{0.5}}
 \put( 59.80, 55.84){\circle*{0.4}}
 \put( 59.90, 59.14){\circle*{0.5}}
 \put( 59.90, 55.89){\circle*{0.4}}
 \put( 60.00, 59.19){\circle*{0.5}}
 \put( 60.00, 55.93){\circle*{0.4}}
 \put( 60.10, 59.24){\circle*{0.5}}
 \put( 60.10, 55.98){\circle*{0.4}}
 \put( 60.20, 59.29){\circle*{0.5}}
 \put( 60.20, 56.02){\circle*{0.4}}
 \put( 60.30, 59.35){\circle*{0.5}}
 \put( 60.30, 56.07){\circle*{0.4}}
 \put( 60.40, 59.40){\circle*{0.5}}
 \put( 60.40, 56.11){\circle*{0.4}}
 \put( 60.50, 59.45){\circle*{0.5}}
 \put( 60.50, 56.16){\circle*{0.4}}
 \put( 60.60, 59.50){\circle*{0.5}}
 \put( 60.60, 56.20){\circle*{0.4}}
 \put( 60.70, 59.55){\circle*{0.5}}
 \put( 60.70, 56.24){\circle*{0.4}}
 \put( 60.80, 59.61){\circle*{0.5}}
 \put( 60.80, 56.29){\circle*{0.4}}
 \put( 60.90, 59.66){\circle*{0.5}}
 \put( 60.90, 56.33){\circle*{0.4}}
 \put( 61.00, 59.71){\circle*{0.5}}
 \put( 61.00, 56.38){\circle*{0.4}}
 \put( 61.10, 59.76){\circle*{0.5}}
 \put( 61.10, 56.42){\circle*{0.4}}
 \put( 61.20, 59.81){\circle*{0.5}}
 \put( 61.20, 56.47){\circle*{0.4}}
 \put( 61.30, 59.86){\circle*{0.5}}
 \put( 61.30, 56.51){\circle*{0.4}}
 \put( 61.40, 59.91){\circle*{0.5}}
 \put( 61.40, 56.55){\circle*{0.4}}
 \put( 61.50, 59.97){\circle*{0.5}}
 \put( 61.50, 56.60){\circle*{0.4}}
 \put( 61.60, 60.02){\circle*{0.5}}
 \put( 61.60, 56.64){\circle*{0.4}}
 \put( 61.70, 60.07){\circle*{0.5}}
 \put( 61.70, 56.68){\circle*{0.4}}
 \put( 61.80, 60.12){\circle*{0.5}}
 \put( 61.80, 56.73){\circle*{0.4}}
 \put( 61.90, 60.17){\circle*{0.5}}
 \put( 61.90, 56.77){\circle*{0.4}}
 \put( 62.00, 60.22){\circle*{0.5}}
 \put( 62.00, 56.81){\circle*{0.4}}
 \put( 62.10, 60.27){\circle*{0.5}}
 \put( 62.10, 56.86){\circle*{0.4}}
 \put( 62.20, 60.32){\circle*{0.5}}
 \put( 62.20, 56.90){\circle*{0.4}}
 \put( 62.30, 60.37){\circle*{0.5}}
 \put( 62.30, 56.94){\circle*{0.4}}
 \put( 62.40, 60.42){\circle*{0.5}}
 \put( 62.40, 56.99){\circle*{0.4}}
 \put( 62.50, 60.47){\circle*{0.5}}
 \put( 62.50, 57.03){\circle*{0.4}}
 \put( 62.60, 60.52){\circle*{0.5}}
 \put( 62.60, 57.07){\circle*{0.4}}
 \put( 62.70, 60.57){\circle*{0.5}}
 \put( 62.70, 57.12){\circle*{0.4}}
 \put( 62.80, 60.62){\circle*{0.5}}
 \put( 62.80, 57.16){\circle*{0.4}}
 \put( 62.90, 60.67){\circle*{0.5}}
 \put( 62.90, 57.20){\circle*{0.4}}
 \put( 63.00, 60.72){\circle*{0.5}}
 \put( 63.00, 57.24){\circle*{0.4}}
 \put( 63.10, 60.77){\circle*{0.5}}
 \put( 63.10, 57.29){\circle*{0.4}}
 \put( 63.20, 60.82){\circle*{0.5}}
 \put( 63.20, 57.33){\circle*{0.4}}
 \put( 63.30, 60.87){\circle*{0.5}}
 \put( 63.30, 57.37){\circle*{0.4}}
 \put( 63.40, 60.92){\circle*{0.5}}
 \put( 63.40, 57.41){\circle*{0.4}}
 \put( 63.50, 60.97){\circle*{0.5}}
 \put( 63.50, 57.46){\circle*{0.4}}
 \put( 63.60, 61.02){\circle*{0.5}}
 \put( 63.60, 57.50){\circle*{0.4}}
 \put( 63.70, 61.07){\circle*{0.5}}
 \put( 63.70, 57.54){\circle*{0.4}}
 \put( 63.80, 61.12){\circle*{0.5}}
 \put( 63.80, 57.58){\circle*{0.4}}
 \put( 63.90, 61.17){\circle*{0.5}}
 \put( 63.90, 57.62){\circle*{0.4}}
 \put( 64.00, 61.21){\circle*{0.5}}
 \put( 64.00, 57.67){\circle*{0.4}}
 \put( 64.10, 61.26){\circle*{0.5}}
 \put( 64.10, 57.71){\circle*{0.4}}
 \put( 64.20, 61.31){\circle*{0.5}}
 \put( 64.20, 57.75){\circle*{0.4}}
 \put( 64.30, 61.36){\circle*{0.5}}
 \put( 64.30, 57.79){\circle*{0.4}}
 \put( 64.40, 61.41){\circle*{0.5}}
 \put( 64.40, 57.83){\circle*{0.4}}
 \put( 64.50, 61.46){\circle*{0.5}}
 \put( 64.50, 57.87){\circle*{0.4}}
 \put( 64.60, 61.51){\circle*{0.5}}
 \put( 64.60, 57.91){\circle*{0.4}}
 \put( 64.70, 61.55){\circle*{0.5}}
 \put( 64.70, 57.96){\circle*{0.4}}
 \put( 64.80, 61.60){\circle*{0.5}}
 \put( 64.80, 58.00){\circle*{0.4}}
 \put( 64.90, 61.65){\circle*{0.5}}
 \put( 64.90, 58.04){\circle*{0.4}}
 \put( 65.00, 61.70){\circle*{0.5}}
 \put( 65.00, 58.08){\circle*{0.4}}
 \put( 65.10, 61.75){\circle*{0.5}}
 \put( 65.10, 58.12){\circle*{0.4}}
 \put( 65.20, 61.80){\circle*{0.5}}
 \put( 65.20, 58.16){\circle*{0.4}}
 \put( 65.30, 61.84){\circle*{0.5}}
 \put( 65.30, 58.20){\circle*{0.4}}
 \put( 65.40, 61.89){\circle*{0.5}}
 \put( 65.40, 58.24){\circle*{0.4}}
 \put( 65.50, 61.94){\circle*{0.5}}
 \put( 65.50, 58.28){\circle*{0.4}}
 \put( 65.60, 61.99){\circle*{0.5}}
 \put( 65.60, 58.32){\circle*{0.4}}
 \put( 65.70, 62.03){\circle*{0.5}}
 \put( 65.70, 58.37){\circle*{0.4}}
 \put( 65.80, 62.08){\circle*{0.5}}
 \put( 65.80, 58.41){\circle*{0.4}}
 \put( 65.90, 62.13){\circle*{0.5}}
 \put( 65.90, 58.45){\circle*{0.4}}
 \put( 66.00, 62.18){\circle*{0.5}}
 \put( 66.00, 58.49){\circle*{0.4}}
 \put( 66.10, 62.22){\circle*{0.5}}
 \put( 66.10, 58.53){\circle*{0.4}}
 \put( 66.20, 62.27){\circle*{0.5}}
 \put( 66.20, 58.57){\circle*{0.4}}
 \put( 66.30, 62.32){\circle*{0.5}}
 \put( 66.30, 58.61){\circle*{0.4}}
 \put( 66.40, 62.36){\circle*{0.5}}
 \put( 66.40, 58.65){\circle*{0.4}}
 \put( 66.50, 62.41){\circle*{0.5}}
 \put( 66.50, 58.69){\circle*{0.4}}
 \put( 66.60, 62.46){\circle*{0.5}}
 \put( 66.60, 58.73){\circle*{0.4}}
 \put( 66.70, 62.51){\circle*{0.5}}
 \put( 66.70, 58.77){\circle*{0.4}}
 \put( 66.80, 62.55){\circle*{0.5}}
 \put( 66.80, 58.81){\circle*{0.4}}
 \put( 66.90, 62.60){\circle*{0.5}}
 \put( 66.90, 58.85){\circle*{0.4}}
 \put( 67.00, 62.65){\circle*{0.5}}
 \put( 67.00, 58.89){\circle*{0.4}}
 \put( 67.10, 62.69){\circle*{0.5}}
 \put( 67.10, 58.93){\circle*{0.4}}
 \put( 67.20, 62.74){\circle*{0.5}}
 \put( 67.20, 58.97){\circle*{0.4}}
 \put( 67.30, 62.78){\circle*{0.5}}
 \put( 67.30, 59.01){\circle*{0.4}}
 \put( 67.40, 62.83){\circle*{0.5}}
 \put( 67.40, 59.04){\circle*{0.4}}
 \put( 67.50, 62.88){\circle*{0.5}}
 \put( 67.50, 59.08){\circle*{0.4}}
 \put( 67.60, 62.92){\circle*{0.5}}
 \put( 67.60, 59.12){\circle*{0.4}}
 \put( 67.70, 62.97){\circle*{0.5}}
 \put( 67.70, 59.16){\circle*{0.4}}
 \put( 67.80, 63.02){\circle*{0.5}}
 \put( 67.80, 59.20){\circle*{0.4}}
 \put( 67.90, 63.06){\circle*{0.5}}
 \put( 67.90, 59.24){\circle*{0.4}}
 \put( 68.00, 63.11){\circle*{0.5}}
 \put( 68.00, 59.28){\circle*{0.4}}
 \put( 68.10, 63.15){\circle*{0.5}}
 \put( 68.10, 59.32){\circle*{0.4}}
 \put( 68.20, 63.20){\circle*{0.5}}
 \put( 68.20, 59.36){\circle*{0.4}}
 \put( 68.30, 63.24){\circle*{0.5}}
 \put( 68.30, 59.40){\circle*{0.4}}
 \put( 68.40, 63.29){\circle*{0.5}}
 \put( 68.40, 59.44){\circle*{0.4}}
 \put( 68.50, 63.34){\circle*{0.5}}
 \put( 68.50, 59.47){\circle*{0.4}}
 \put( 68.60, 63.38){\circle*{0.5}}
 \put( 68.60, 59.51){\circle*{0.4}}
 \put( 68.70, 63.43){\circle*{0.5}}
 \put( 68.70, 59.55){\circle*{0.4}}
 \put( 68.80, 63.47){\circle*{0.5}}
 \put( 68.80, 59.59){\circle*{0.4}}
 \put( 68.90, 63.52){\circle*{0.5}}
 \put( 68.90, 59.63){\circle*{0.4}}
 \put( 69.00, 63.56){\circle*{0.5}}
 \put( 69.00, 59.67){\circle*{0.4}}
 \put( 69.10, 63.61){\circle*{0.5}}
 \put( 69.10, 59.71){\circle*{0.4}}
 \put( 69.20, 63.65){\circle*{0.5}}
 \put( 69.20, 59.74){\circle*{0.4}}
 \put( 69.30, 63.70){\circle*{0.5}}
 \put( 69.30, 59.78){\circle*{0.4}}
 \put( 69.40, 63.74){\circle*{0.5}}
 \put( 69.40, 59.82){\circle*{0.4}}
 \put( 69.50, 63.79){\circle*{0.5}}
 \put( 69.50, 59.86){\circle*{0.4}}
 \put( 69.60, 63.83){\circle*{0.5}}
 \put( 69.60, 59.90){\circle*{0.4}}
 \put( 69.70, 63.88){\circle*{0.5}}
 \put( 69.70, 59.93){\circle*{0.4}}
 \put( 69.80, 63.92){\circle*{0.5}}
 \put( 69.80, 59.97){\circle*{0.4}}
 \put( 69.90, 63.96){\circle*{0.5}}
 \put( 69.90, 60.01){\circle*{0.4}}
 \put( 70.00, 64.01){\circle*{0.5}}
 \put( 70.00, 60.05){\circle*{0.4}}

\put(71.5,64){\small 1}
\put(71.5,58.5){\small 2}

\end{picture}
\end{center}
\caption{\small The experimental values of the $nS$ bottomonium (solid dots)
and charmonium (empty boxes) mass differences 
$\alpha(n) =[M(nS)-M(1S)]/[M(2S)-M(1S)]$
and the dependences in the present model (curve  1) and in 
the WKB approximation (curve  2).}
\label{fm}
\end{figure}
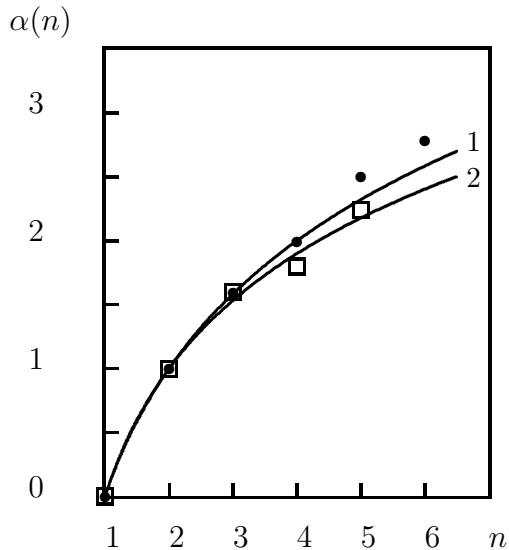
The mass of $nS$-level is determined by the following
$$
M(nS) = m_1+m_2+ E(n)\;.
$$
For the sake of convenience we introduce the "initial"
value $n_i(\mu_{12})$ depending on the reduced mass and related to the 
flavour-independent constant $C$ in (\ref{4})
\begin{equation}
C= - T\ln{\frac{n_i^2(\mu_{12})}{\mu_{12}}}\;. \label{6}
\end{equation}
Let us use the experimental data on the heavy quarkonium spectra
$$
M_\Upsilon(4S) \approx 2m_B(1S)\;,\;\;\;
M_\psi(3S) \approx 2m_D(1S)\;,
$$
which are valid with the 30 MeV accuracy. These equations can be 
rewritten down as\footnote{The analogous 
estimate with the additional assumption $n_i(b\bar b) = 1$ was considered in
\cite{10-}.}
\begin{eqnarray}
2 T \ln{\frac{n_{th}(b\bar b)}{n_i(b\bar b)}} & = & 
2\bar \Lambda +\frac{\mu^2_\pi}{m_b}\;, \label{7}\\
2 T \ln{\frac{n_{th}(c\bar c)}{n_i(c\bar c)}} & = & 
2\bar \Lambda +\frac{\mu^2_\pi}{m_c}\;,\label{8}
\end{eqnarray}
where $n_{th}(b\bar b) = 4$, $n_{th}(c\bar c) = 3$. From eq.(\ref{6})
one can find
\begin{equation}
\ln{\frac{n_i(b\bar b)}{n_i(c\bar c)}} = \frac{1}{2}\;
\ln{\frac{m_b}{m_c}}\;. \label{9}
\end{equation}
Combining (\ref{2}), (\ref{7}-\ref{9}), one gets
\begin{equation}
\bar \Lambda = \frac{m_b m_c}{m_b-m_c} \ln\bigg\{\sqrt{\frac{m_b}{m_c}}
\frac{n_{th}(c\bar c)}{n_{th}(b\bar b)}\bigg\}\;, \label{10}
\end{equation}
where the dependence on the $T$ parameter is hidden in explicit relations 
for the quark masses through the heavy-light meson masses and $\bar \Lambda$.

Eq.(\ref{10}) can be solved numerically, and it gives\footnote{The obtained 
result is in a good agreement with the restrictions derived in \cite{10=}.}
\begin{equation}
\bar \Lambda = 0.63\pm 0.03\; \mbox{GeV.}
\end{equation}
As for the quark masses in the first order over $1/m_Q$, one finds
$$
m_b=4.63\pm 0.03\; \mbox{GeV,}\;\;\;
m_c=1.18\pm 0.07\; \mbox{GeV.}
$$
The additional uncertainty in $c$ quark mass estimate is related to the 
replacement $\bar \Lambda/m(Q\bar q) \to \bar \Lambda/(m(Q\bar q)-\bar \Lambda)
+O(\bar \Lambda^2/m^2_Q)$ in the heavy quark mass expression, i.e.
it is caused by terms of the second order over the inverse heavy quark mass.

The $\mu^2_\pi$ parameter is equal to $0.54\pm 0.08$ GeV$^2$ \cite{1,9+,10+}.

The performed calculations allow one to predict the $nS$-level masses of
$\bar b c$ family below the $BD$ pair threshold
$$
m_{B_c}(1S) =6.37\pm 0.04\; \mbox{GeV,}\;\;\;
m_{B_c}(2S) =6.97\pm 0.04\; \mbox{GeV.}
$$
The $1S$-level position is slightly higher than in previous estimates 
in the framework of potential models \cite{11}. This deviation is 
basically caused by the greater value of $T$ parameter, but not by
the difference in the quark mass values, since the $\bar b c$ level masses are
not very much sensitive to the quark mass variation. Using the estimate
of spin-dependent splitting of $1S$-level in $\bar b c$ system,
$m(1^-)-m(0^-)\approx 60 -70$ MeV \cite{11}, one gets the mass of
the basic pseudoscalar state
$$
m_{B_c}(0^-) = 6.32\pm 0.05\; \mbox{GeV.}
$$

\section*{Conclusion}

Using the regularity of heavy quarkonium spectra described within the
quasiclassical approach, we have evaluated the heavy quark-meson mass gap
$\bar \Lambda = 0.63\pm 0.03$ GeV, as well as the pole masses of $b$ and $c$
quarks. The $B_c$ meson mass can be predicted $m_{B_c} =6.32\pm 0.05$ GeV.

\section*{Acknowledgement}

This work is in part supported by the International Science Foundation
grants NJQ000, NJQ300 and by the program "Russian State Stipends for Young
Scientists".
The author thanks prof. N.~Paver and O.~Yushchenko for 
stimulating discussions.

\vspace*{4mm}

\hfill {\it Received October 19, 1995}
\end{document}